\documentstyle[prl,aps,epsf,twocolumn]{revtex}

\begin{document}

\draft
\title{Tunneling problems by quantum Monte Carlo}
\author{Nikolai  Prokof'ev, Boris  Svistunov, and Igor Tupitsyn}
\address{Russian Research Center ``Kurchatov Institute",
123182 Moscow, Russia}
\maketitle

\begin{abstract}
We develop a new numerical scheme which allows precise solution
of coherent tunneling problems, i.e., problems with exponentially small 
transition amplitudes between quasidegenerate states. We explain
how this method works for the single-particle (tunneling in the
double-well potential) and many-body systems (e.g.,
vacuum-to-vacuum transitions),  and gives directly the
instanton shape and tunneling amplitude. Most importantly, 
transition amplitudes may be calculated to arbitrary accuracy
(being limited solely by statistical errors) no matter 
how small are their absolute values.
\end{abstract}



Tunneling phenomena are among the most intriguing consequences
of quantum theory. They are of fundamental importance both for
the high-energy and condensed matter physics, and the list of
systems which behavior is governed by tunneling transitions
ranges from quantum chromodynamics \cite{Belavin,Shifman} to
Josephson junctions and defects in crystals (see, e.g.,
Ref.~\onlinecite{book}). 

Precise analytic treatment of tunneling in complex systems is
very hard, if not impossible. The most crucial simplification is
in reducing the original problem to the semiclassical study of
the effective action for some collective variable 
${\bf R}$, with the assumption that all the other degrees of
freedom adjust adiabatically to the motion of ${\bf R}$ 
\cite{Lifshitz,book}. In certain cases, one may also include
dissipative effects due to ``slow'' modes other than the selected
collective variable which {\it do not} follow the dynamics of 
${\bf R}$ adiabatically \cite{book,Weiss}. Typically, the
parameters of the effective  action can not be found
analytically (although one may relate some of them to the linear
response coefficients) and have to be deduced from experiments.
As far as we are aware, at present there are no tools to address
the tunneling problem numerically. ``Exact'' diagonalization 
works only for relatively small systems, and, even in small
systems, its accuracy is not sufficient to resolve very
small energy splittings $\Delta E$, 
say, when $\Delta E/E \ll 10^{-10}$, due to
round-off errors (unless specific no-round-off arithmetics is used).

In this letter we develop a quantum Monte Carlo (MC) approach
which allows precise calculations of tunneling amplitudes (and
instanton shapes) no matter how small are their absolute values. The
MC scheme contains no systematic errors, and its accuracy
is limited only by statistical noise. The key point of our
approach is in simulating imaginary-time dependence of 
transition amplitudes $A_{ji}(\tau )$ between selected
reference states $\vert \eta_1 \rangle$ and $\vert \eta_2 \rangle$
(see below). In doing so we have to solve the problem of
collecting reliable statistics in a case when $A_{ji}(\tau )$
varies, say,  over hundreds of
orders of magnitude (!) between different points in time. 
First, we describe in detail how to evaluate tunneling
splitting and instanton shape in the double-well potential. We
proceed then to the many-body problem of vacuum-to-vacuum
transitions by considering the  case of 1D quantum antiferromagnet
with exchange anysotropy. Finally, we
discuss the generality of the method suggested and 
demonstrate our numeric results for tunneling in the
double-well potential (with comparison to the 
exact-diagonalization data where possible).

Consider the standard problem of particle motion in
external potential: 
\begin{equation}
H=m\dot{x}^2/2 +U(x) \;,
\label{H-DW}
\end{equation}
which has two minima at points $x=\eta_1$ and $x=\eta_2$, and
large tunneling action $S=\int^{\eta_2}_{\eta_1} p \, dx =
\int^{\eta_2}_{\eta_1} [2mU(x)]^{1/2} dx \gg 1$. These minima
are supposed to be near-degenerate, i.e., the lowest eigenstates
of the Hamiltonian (\ref{H-DW}), $H \Psi_{\alpha}
=E_{\alpha}\Psi_{\alpha}$, form a doublet with
\begin{equation}
E_2-E_1 \sim e^{-S}\omega_{i } \ll \omega_{i } \;,
\label{inequality}
\end{equation}
where $\omega_{i}$ are the classical vibration frequencies in the
potential minima $\omega_{i } =[U''(\eta_i )m]^{1/2}$ 
(we set $\hbar =1$). 
The lattice analog of the Hamiltonian (\ref{H-DW}) reads
\begin{equation}
H=-t\sum_{<ll'>} d^{\dag }_ld_{l'} + \sum_{l} n_l U_l \;, \;\;\;\;
n_l =d^{\dag }_ld_l \;,
\label{H-DW-L}
\end{equation}
where $d^{\dag }_l$ creates a particle on the site number $l$, 
and the first sum is over nearest-neighbor sites.

The transition amplitude from the state $\vert \eta_1 \rangle$ 
to the state  $\vert \eta_2 \rangle$, where
$\vert \eta_{1,2} \rangle=\delta (x-\eta_{1,2} )$,
in imaginary time $\tau$ is given by
\begin{equation}
A_{ji}(\tau ) = \langle \eta_2 \vert e^{-H\tau }\vert \eta_1 \rangle \equiv 
\sum_{\alpha } \langle \alpha \vert \eta_1 \rangle 
\langle \eta_2 \vert \alpha \rangle e^{-E_{\alpha} \tau } \;.
\label{Ampl}
\end{equation}
We now make use of the inequality (\ref{inequality}) to define
the asymptotic regime $E_2-E_1 \ll \tau^{-1} \ll \omega_{i} $,
see, e.g., Refs.~\onlinecite{Coleman},
\begin{equation}
A_{ji}(\tau ) \to e^{-\bar{E} \tau } \sum_{\alpha =1,2} 
\langle \alpha \vert \eta_1 \rangle 
\langle \eta_2 \vert \alpha \rangle [1-(E_{\alpha}-\bar{E})\tau
] \;. 
\label{Ampl-long}
\end{equation}
where $\bar{E}=(E_2+E_1)/2$.

It is convenient to split the double-well potential in two terms
$U(x)=U^{(1)} (x)+ U^{(2)} (x)$, where
$U^{(1,2)} (x)$ is identical to $U(x)$   
to the left/right of the
barrier maximum point and remains constant afterwards. 
Introducing system ground states in each minimum as
$H^{(i)} \Psi_G^{(i)} = E_G^{(i)} \Psi_G^{(i)}$,
where $H^{(i)}= m\dot{x}^2/2 +U^{(i)}(x)$, we may rewrite
\begin{eqnarray}
\Psi_1 & =& u \Psi_G^{(1)}+v\Psi_G^{(2)} \; ,\nonumber \\
\Psi_2 & =& v \Psi_G^{(1)}-u\Psi_G^{(2)}  \; ,
\label{Psi-Psi}
\end{eqnarray}
where $(u^2,v^2) = 1/2\pm \xi /2E$, $E^2=\Delta^2+\xi^2$, with
obvious identification of the energy splitting $2E=E_2-E_1$ and 
bias energy $\xi = (E_G^{(1)}-E_G^{(2)})/2$. 
Here $\Delta$ is the tunneling
amplitude, which defines energy splitting $E_2-E_1=2\Delta$ in the
degenerate case. Substituting Eq.~(\ref{Psi-Psi}) 
into Eq.~(\ref{Ampl-long}) we finally obtain
\begin{eqnarray}
A_{ii} (\tau ) & \to & e^{-\bar{E} \tau } Z_i^2 \;,\nonumber \\
A_{j\ne i} (\tau ) & \to & \tau e^{-\bar{E} \tau } Z_i Z_j \Delta   \; .
\label{Ampl-fin}
\end{eqnarray}
Here $Z_i=\langle \Psi_G^{(i)} \vert \eta_i \rangle $
projects the reference states $\vert \eta_i \rangle $ on the
corresponding ground states in each minimum. All corrections to
Eq.~(\ref{Ampl-fin}), e.g., the neglect of the overlap integrals
$\langle \Psi_G^{(1)} \vert \eta_2 \rangle $ and
$\langle \Psi_G^{(2)} \vert \eta_1 \rangle $,
 are small in parameter $e^{-S}$. Note also that 
Eq.~(\ref{Ampl-fin}) does not depend on the bias $\xi \ll
\omega_i$, and is insensitive to the behavior of
$\Psi_G^{(i)}$ in the deep underbarrier region.
This (rather standard) consideration relates transition amplitudes to the
tunneling amplitude through the asymptotic analysis of $A_{ij}(\tau )$ in 
time. 


MC simulation of the transition amplitude is almost identical to
the standard simulation of quantum statistics [the partition
function at a given temperature $T=1/\beta$ may be written as ${\cal
Z}(\tau =\beta )={\rm Tr}_{\eta } A(\eta ,\eta , \tau )$]. Fixed boundary
conditions, as opposed to the trace over closed trajectories
(configurations), are trivial to deal within any scheme. The
real difference is in sampling different time-scales -
thermodynamic calculations are typically done with $\beta = const$.
Now we are forced to consider
trajectories with different values of $\tau$
and to treat imaginary time as ``dynamic'' (in MC sense)
variable. The idea of utilizing time dependencies of
trajectories in MC simulations was extensively discussed in
connection with the Worm algorithm \cite{JETP,PLA} and polaron
Green function \cite{polaron}.

We now turn to the problem of normalization. This problem might seem
intractable in view of close analogy between ${\cal Z}(\tau )$ and 
$A(\tau )$. Formally, in the limit
$\tau \to 0$, the amplitude is trivial to find in most cases,
e.g., for the Hamiltonians (\ref{H-DW}) and (\ref{H-DW-L}) it is
given by the free particle propagation
\begin{equation}
\!\!\! A_{ji}(\tau \to 0 ) \to \!
\left\{ 
\begin{array}{ll}
e^{-m(\eta_i-\eta_j)^2/2\tau } \sqrt{{m \over 2\pi \tau }} 
& \;\;{\rm continuous} \\
(t\tau )^{\vert \eta_i-\eta_j \vert }/ \vert \eta_i-\eta_j \vert !
& \;\;{\rm discrete} \;, 
\end{array} \right.
\label{A-short}
\end{equation}
and this knowledge may be used to normalize MC statistics for
$A_{ji}(\tau )$ (in the discrete case $\vert \eta_i-\eta_j \vert =$integer).
However the absolute values of 
$A_{ji} (\tau )$ at short and long times will typically differ
by orders and orders of magnitude and simply none MC statistics
will be available at short times. 

The solution to the puzzle lays in the possibility to use an arbitrary
fictitious potential $A_{\mbox{\scriptsize fic}}(\tau )$ in Metropolis-type updates
\cite{Metropolis} in time-domain
\begin{equation}
{P_{acc}({\cal B} \to {\cal A})  \over P_{acc}({\cal A} \to {\cal B})} =
e^{-S_{\cal A}+S_{\cal B}} \; 
{A_{\mbox{\scriptsize fic}} (\tau ') \over A_{\mbox{\scriptsize fic}} (\tau ) }
{W_{\cal A} (\tau ) \over W_{\cal B} (\tau ') } \;,
\label{Metro}
\end{equation}
where $P_{acc}({\cal B} \to {\cal A})/P_{acc}({\cal A} \to {\cal
B})$ is the acceptance ratio for the update transforming
initial trajectory ${\cal B}$ (having duration $\tau $) to the 
trajectory ${\cal A}$ (having duration $\tau '$), and $e^{-S_{\cal B}}$
is the statistical weight of the trajectory ${\cal B}$. 
The normalized distribution functions $W_{\cal B} (\tau ') $,
according to which a new value of $\tau $ is seeded, are also arbitrary;
the best choice of $W$'s follows from the conditions of (i)
optimal acceptance ratio (as close to unity as possible), and
(ii) simple analytic form allowing trivial solution of the equation
$\int_{\tau_a }^{\tau '} W(\tau ') d \tau ' = 
r\int_{\tau_a }^{\tau_b} W(\tau ') d \tau '$ on the time interval
$(\tau_a,\tau_b)$ , where $0<r<1$ is the
random number \cite{JETP,polaron}. Each trajectory adds a contribution
$=1/A_{\mbox{\scriptsize fic}}(\tau )$ to the time-histogram of $A_{ij} (\tau )$.

One may use fictitious potential to enhance statistics of trajectories
with certain values of $\tau$ ``by hand'', e.g., by making $A_{\mbox{\scriptsize fic}}$
zero outside some time-window. To get a reliable and properly weighted
statistics both at short and long times we need $A_{\mbox{\scriptsize fic}} (\tau )
\sim 1/A_{ij} (\tau ) $ to compensate completely for the severe variation
of the transition amplitude between different time-scales. This
goal is achieved as follows. The initial stage of the calculation, 
called thermolization, prepares the fictitious potential using recursive
self-adjusting scheme - starting from $A_{\mbox{\scriptsize fic}} (\tau )=1$ in a given 
time-window $(\tau_{min}, \tau_{max}) $ and zero otherwise, we collect
statistics for $A_{ij} (\tau )$ to the temporary time-histogram and after
$M>10^6\div 10^{7}$ updates we renew the fictitious potential as
\begin{equation}
A_{\mbox{\scriptsize fic}} (\tau )= \left\{
\begin{array}{ll}
A_{ij}( \tau_0 )/A_{ij} (\tau ) \;, & \;\; \tau_1 < \tau < \tau_2 \\
A_{ij}( \tau_0 )/A_{ij} (\tau_1 ) \;, & \;\; \tau_{min} < \tau < \tau_1 \\
A_{ij}( \tau_0 )/A_{ij} (\tau_2 ) \;, & \;\; \tau_2 < \tau < \tau_{max} \\
\end{array} \right.
\label{scheme}
\end{equation}
where $\tau_1$ and $\tau_2$ are the  points (to the left and
to the right of some reference point $\tau_0$) where temporary statistics
becomes unreliable and has large fluctuations. It makes sense to
select $\tau_0$ close to the maximum of $A_{ij}(\tau
)$ (this point may be tuned a posteriory), 
and points $\tau_1 $ and $\tau_2$ are formally defined as
the first points in the histogram where smooth variation of
$A_{ij}(\tau )$ ends:
$A_{ij} (\tau_{1,2}+\Delta \tau )/ A_{ij} (\tau_{1,2} ) < \delta $ (here 
$\Delta \tau$ is the difference between the nearest points
in time-histogram, and $\delta $ is a small number, say 0.01).
The thermolization stage continues until $\tau_1=\tau_{min}$ and
$\tau_2 =\tau_{max} $, and fictitious potential stops changing 
(withing a factor of two). After that the actual calculation starts
with a fixed $A_{\mbox{\scriptsize fic}}(\tau )$, and a new histogram 
for $A_{ij}(\tau )$ is collected. 

The idea of using fictitious potential proportional to the inverse of the
transition amplitude is clear - it allows to collect reliable
statistics on different time-scales with comparable relative accuracy.
With this tool at hand one can easily normalize $A_{ij} (\tau )$
using known analytic results for short times, and deduce
tunneling amplitude and Z-factors from the analysis of 
the long-time asymptotic, Eq.~(\ref{Ampl-fin}). We can not but note that
fictitious potential in time-domain very much resembles the so-called
``guiding wavefunction" in the 
Green-function MC methods \cite{Ceperley}, with essential difference 
that here it is used to reach exponentially rare configurations.  

Instanton shape calculation is a much easier task since it can
be done by considering trajectories for the transition
amplitude $A_{i \ne j} (\tau ) $  with fixed but
sufficiently long $\tau \gg 1/\omega_i$; again, to ensure that
$A_{ij}$ is dominated by just one instanton trajectory we need
$\tau  \ll 1/\Delta$. For any given MC trajectory $x(\tau )$ 
one has first  to define the instanton center position in time,
and to recount all times from this center-point $\tau_c$ 
[instanton center statistics is almost uniform in $(0,\tau )$
(except near the ends of the time interval) due to the generic 
``zero-mode'' present in instanton solutions
 \cite{Shifman,Coleman}].
This can be done by looking at the average time 
\begin{equation}
\tau_c={ \int_B \tau d\tau \over \int_B d\tau} \; ,
\label{zero}
\end{equation}
where the integral is taken over the barrier region between the wells
$U(x(\tau )) > E_G$. The instanton shape is obtained then by
collecting statistics of $x(\tau -\tau_c)$ to the time histogram.
In this simple example when the notion of collective coordinate
is not necessary (or formally, ${\bf R}=x$), we do not need to 
define separately the estimator for ${\bf R}$ (see the opposite
example below).
  
The case of vacuum-to-vacuum transition in the
many-body problem is formulated in precisely the same manner, and 
Eqs.~(\ref{Ampl}-\ref{Ampl-fin}), (\ref{Metro}-\ref{scheme})
holds true once the identification of the  states
$\vert \eta_{1,2} \rangle $ is done and the short-time limit,
Eq.~(\ref{A-short}), is calculated. Consider as a typical
example a 1D spin chain with $2L$ sites and antiferromagnetic
(AF) couplings between the nearest-neighbor spins
\begin{equation}
H=  \sum_{<ij>} \left[ J \vec{S}_i \vec{S}_j + 
J' S_i^z S_j^z  \right]  \;,
\label{AF}
\end{equation}
which has near degenerate ground state with the lowest doublet
separated from the rest of the system spectrum by finite gap for
$J' > 0$. The natural choice of $\vert \eta_{1,2} \rangle $
is then an ordered AF state with $S_i^z \vert \eta_{1,2} \rangle =
(\pm 1)^i \vert \eta_{1,2} \rangle$. Note, that for a large system
Z-factors $\langle \Psi^{(i)}_G \vert \eta_{i} \rangle$
will be also exponentially small, and even diagonal amplitudes
$A_{ii}$ have to be calculated with the use of $A_{\mbox{\scriptsize fic}}$ [in the
single-particle case one may obtain Z-factors by ignoring $A_{\mbox{\scriptsize fic}}$-trick].
The short-time behavior is given by 
\begin{eqnarray}
A_{ii}(\tau )  &\to &  1  \nonumber \\
A_{i\ne j}(\tau ) & \to & \left( J\tau /2\right) ^L 
\left\{  \begin{array}{ll}
2 & \;\;{\rm (ring)} \\
1 & \;\;{\rm (open \;\; chain)} \;,    \end{array} \right.
\label{A-short-AF}
\end{eqnarray}

To decipher the instanton we need now some {\it a priori} knowledge 
about the relevant collective variable (if such knowledge is not
available one has to study different possibilities). For example,
if tunneling proceeds via two domain walls well-separated from
each other (thin-wall approximation \cite{Lifshitz,Shifman,Coleman}),
then the collective variable ${\bf R}$ is the distance between 
the walls, and the underbarrier region in Eq.~(\ref{zero})
is related to the existence of two separated walls. 
These definitions are not too specific and work only
approximately; this is however the generic difficulty of dealing
with collective variables which are meaningful only in the
macroscopic limit. Obviously, the knowledge of the instanton shape does not
allow precise evaluation of $\Delta $, and gives
only rough estimation of $\ln \Delta $.

To test the proposed scheme and to compare results to the
exact diagonalization (ED) data we have applied our algorithm to the lattice 
model (\ref{H-DW-L}) with $U(x)= U_0[(x/\eta )^2-1]^2$. In what follows we 
measure all energies in units of the hopping amplitude $t$ and
count them from the potential minimum. We
set $U_0=1$ and consider two interwell separations: $\eta =10 $ 
and $\eta =40$. For $\eta = 10$ the ED data for the ground state
energy, Z-factor, and tunneling amplitude are: $E_G=(E_1+E_2)/2=0.1923$,
$Z^2=0.2465$, $\Delta = (E_2-E_1)/2 =3.6078 \times 10^{-6}$. Our
MC data give $E_G=0.192(2)$, $Z^2=0.246(2)$, and 
$\Delta =3.61(3) \times 10^{-6}$. The case of large $\eta $
is more subtle since only $E_G=0.0495$ and $Z^2=0.1255$ may be
tested against ED - one may use textbook semiclassical
analysis of the corresponding continuous model to see that
$\Delta (\eta =40) \sim 10^{-24}\div 10^{-23}$,   
that is far beyond the standard computer round-off
errors. 

\vspace{-3cm}
\begin{figure}
\begin{center}
\leavevmode
\epsfxsize=0.40\textwidth
\epsfbox{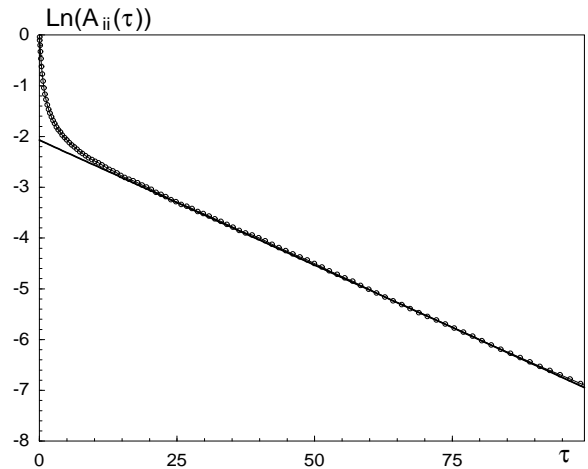}
\end{center}
\caption{Time dependence of the diagonal amplitude $A_{ii}$. 
Solid line $Z^2e^{-E_G \tau }$ is the prediction of the long-time 
behavior according to the ED data.}
\label{fig:fig2}
\end{figure}

\begin{figure}
\begin{center}
\leavevmode
\epsfxsize=0.40\textwidth
\epsfbox{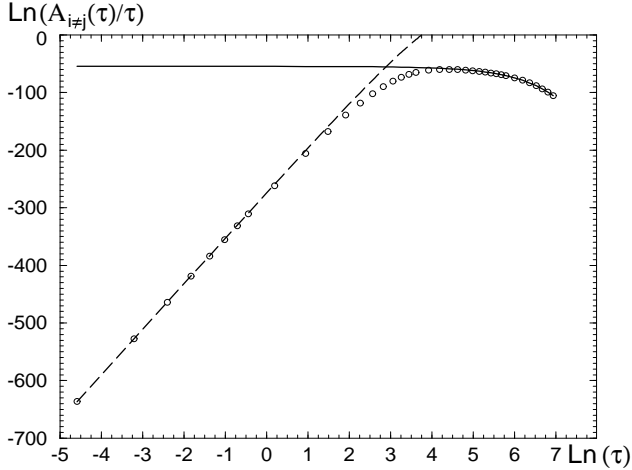}
\end{center}
\caption{Time dependence of the non-diagonal normalized amplitude 
$A_{i\ne j}$.
The solid line is fitting to the curve $\Delta Z^2 e^{-E_G\, \tau }$
and the short-time behavior is tested against the law 
$ (\tau^{2\eta-1} / (2\eta)!) e^{ -\bar{U} \tau}$ (dashed curve).}
\label{fig:fig3}
\end{figure}

In Fig.~(\ref{fig:fig2}) and Fig.~(\ref{fig:fig3})
 we present our MC data for the
diagonal and off-diagonal amplitudes for $\eta=40$, and fits to the 
expected long-time,
Eq.~(\ref{Ampl-fin}), and short-time behavior, Eq.~(\ref{A-short}) [we
also include the lowest-order correction for the potential energy
at $\tau \to 0$ which tells that $A_{i \ne j} \propto e^{-\bar{U}
\tau }$ where $\bar{U}=\int_{-\eta}^{\eta} U(x) dx$]. The
variation of $A_{i \ne j}$ in Fig.~2 is about two-hundred orders!
From Fig.~(\ref{fig:fig2}) 
we deduced $E_G=0.0494(2)$ and $Z^2=0.126(1)$ in agreement
with ED. From Fig.~(\ref{fig:fig3}) we then obtain 
$\Delta = 1.7(4) \times 10^{-23}$. MC simulation of the above
figures took about 5 days each on PII-266 processor.

\vspace{0.7cm}

\vspace{0.0cm}
\begin{figure}
\begin{center}
\leavevmode
\epsfxsize=0.40\textwidth
\epsfbox{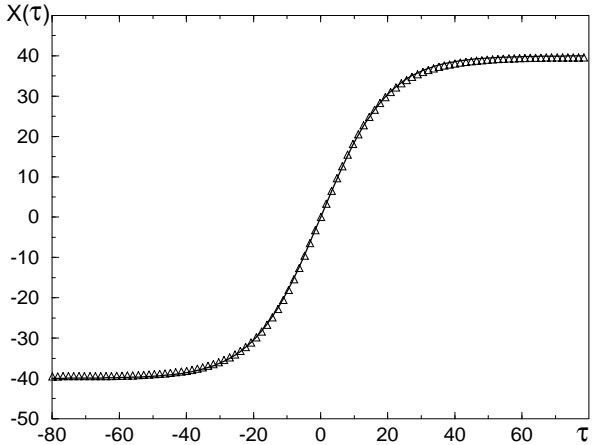}
\end{center}
\caption{Instanton trajectory $x(\tau )$. The solid curve
$x(\tau ) = \eta \tanh (E_G \, \tau )$ is the semiclassical result
for the continuous model.}
\label{fig:fig4}
\end{figure}

In Fig.~(\ref{fig:fig4}) we present our results for the
instanton trajectory $x(\tau )$ (at $\eta =40$)
along with the known semiclassical results for the continuous
model (\ref{H-DW}) \cite{Shifman}. The accuracy of the data is
self-evident, 
although we argue that MC data rather represent the trajectory
with finite energy $E=E_G$ while analytic results correspond to $E=0$.   

We note that the present technique is very hard
to implement in the discrete-time schemes with finite Trotter
parameter $\Delta \tau$. On one hand, long-time asymptotic
regime (see Fig.~2) requires to consider $\tau $ as long as $640$. 
On another hand at short times, the requirement of smooth variation 
of the amplitude 
$\Delta \tau \, d\ln [A_{i\ne j}(\tau )]/d\tau \ll 1$, after
substituting Eq.~(\ref{A-short}), means $\Delta \tau \ll \tau /(2 \eta )
\approx 10^{-3}$ for $\tau =0.1$. To avoid large systematic errors
due to time-discretization at short times one has
to use $\Delta \tau =10^{-4}$(!!). Apart from enormous
memory usage (there will be about $10^7$ time slices) 
that small Trotter parameter severely slows down the efficiency 
of the code, in fact, we are not aware of any MC simulation with
$\Delta \tau \sim 10^{-4}$.

It is worth mentioning that similar technique makes it possible
to study directly ${\cal Z}(\tau )$ over different time-scales -
normalization of the partition function does not matter since none
physical quantity depend on it. Thus one can obtain temperature dependences
of the free energy, entropy etc. in a single MC run.

This work was supported by the RFBR Grants 
98-02-16262 and 97-02-16548 (Russia),   
IR-97-2124 (European Community).


\end{document}